\documentclass[conference]{IEEEtran}
\IEEEoverridecommandlockouts
\usepackage{cite}
\usepackage{amsmath,amssymb,amsfonts}
\usepackage{graphicx}
\usepackage{xcolor}
\def\BibTeX{{\rm B\kern-.05em{\sc i\kern-.025em b}\kern-.08em
    T\kern-.1667em\lower.7ex\hbox{E}\kern-.125emX}}
\pdfoutput=1
\usepackage{cite}
\ifCLASSINFOpdf
\else
\fi
\usepackage{stix}
\newcommand{\field}[1]{\mathbb{#1}}

\newcommand{\R}{\field{R}} 
\newcommand{\commentOut}[1]{}
\newcommand{\Th}{^{\text{th}}}

\usepackage{mathrsfs}
\usepackage{subcaption}
\usepackage{svg}
\usepackage{amsmath,amssymb,amsthm}
\usepackage{algorithm}
\usepackage{comment}
\usepackage{float}
\usepackage{array}
\usepackage{algpseudocode}
\usepackage{makecell}
\usepackage{graphicx}
\usepackage{multirow}
\usepackage{balance}
\usepackage{dirtytalk}

\makeatletter
\newcommand*\titleheader[1]{\gdef\@titleheader{#1}}
\AtBeginDocument{%
\let\st@red@title\@title
\def\@title{%
\bgroup\normalfont\large\centering\@titleheader\par\egroup
\vskip1.5em\st@red@title}
}
\makeatother

\makeatletter
\def\endthebibliography{%
	\def\@noitemerr{\@latex@warning{Empty `thebibliography' environment}}%
	\endlist
}
\makeatother

\begin{document}
\title{An Adversarial Approach to Evaluating the Robustness of Event Identification Models\\
}

\author{\IEEEauthorblockN{\IEEEauthorblockA{\text{Obai~Bahwal \IEEEmembership{Member,~IEEE,}} \text{Oliver~Kosut \IEEEmembership{Senior~Member,~IEEE}}, \text{and~Lalitha~Sankar \IEEEmembership{Senior~Member,~IEEE}}}
\text{School of Electrical, Computer, and Energy
			Engineering, Arizona State University, Tempe, USA} \\
\{obahwal,lalithasankar,okosut\}@asu.edu}

    \thanks{The work of Kosut and Sankar is funded in part by the NSF EPCN-2246658 and a DoE BIRD grant. Bahwal is funded by a grant from the Saudi Arabian Cultural Mission in the United States.}

}

	\markboth{IEEE TRANSACTIONS ON POWER SYSTEMS}%
	{O.Bahwal\MakeLowercase{\textit{et al.}}: An Adversarial Approach to Evaluating the Robustness of Event Identification Models}
	
\maketitle

\begin{abstract}

Intelligent machine learning approaches are finding active use for event detection and identification that allow real-time situational awareness. Yet, such machine learning algorithms have been shown to be susceptible to adversarial attacks on the incoming telemetry data. 
This paper considers a physics-based modal decomposition method to extract features for event classification and focuses on interpretable classifiers including logistic regression and gradient boosting to distinguish two types of events: load loss and generation loss. The resulting classifiers are then tested against an adversarial algorithm to evaluate their robustness. The adversarial attack is tested in two settings: the white box setting, wherein the attacker knows exactly the classification model; and the gray box setting, wherein the attacker has access to historical data from the same network as was used to train the classifier, but does not know the classification model.  Thorough experiments on the synthetic South Carolina 500-bus system highlight that a relatively simpler model such as logistic regression is more susceptible to adversarial attacks than gradient boosting.
%
\end{abstract}



\IEEEpeerreviewmaketitle
\section{Introduction}
\IEEEPARstart{W}{ith} the increasing need for real-time monitoring of the grid dynamics, machine learning (ML) algorithms are providing viable and highly accurate solutions that support the system operator's requirement in making informed and timely decisions for reliable and safe operation of the system. In particular, such algorithms are invaluable for leveraging high-fidelity synchrophasor data (obtained using phasor measurement units (PMUs)) in real-time for 
accurate event detection \cite{realtime, ems}. However, PMUs have been shown to be susceptible to adversarial attacks~\cite{fdi1, fdi2} which in turn can lead to erroneous outcomes from the learned ML models.

In \cite{fdi3}, the authors evaluate false data injection (FDI) attacks on ML-based state estimation models that rely on supervisory control and data acquisition (SCADA) network. The authors use data poisoning and gradient-based attacks as the threat models and show that such attacks are very successful in causing the state estimator to fail. 
More recently, \cite{dnn} evaluates white box adversarial attacks against event classification models based on deep neural networks. Those models utilize time-series PMU measurements to classify between `no events', `voltage-related', `frequency-related', or `oscillation-related' events. 

In contrast to the above-mentioned recent results, \emph{we focus on real-time event identification using PMU data and physics-based modal decomposition methods along with interpretable ML models}. Our event identification framework leverages the approach in \cite{N-et-al-2023} and involves two steps: (i) extract features using physics-based modal decomposition methods; (ii) use such features to learn logistic regression (LR) and gradient boosting (GB) models for event classification. \emph{Our primary goal is to design an algorithmic approach that generates adversarial examples to evaluate the robustness of this physics-based event classification framework}. We evaluate our attack algorithm in two distinct settings: white box and gray box. In the white box setup, we assume that the attacker has full knowledge of the classification framework including the classification model  (i.e., knows both (i) and (ii) detailed above), and can only tamper with a subset of PMUs. On the other hand, for the gray box setup, we assume that the attacker does not know the ML classifier used by the system operator or the data that was used for training; however, the attacker has knowledge of the aspect (i) of the framework, has access to historical data from the same network, and can tamper with a subset of PMUs. In either setting, the attack algorithm perturbs event features in the direction of the classifier's gradient until the event is incorrectly classified. Using detailed event-inclusive PSS/E generated synthetic data for the 500-bus South Carolina system, we show that both types of attacks can significantly reduce the accuracy of the event classification framework presented in \cite{N-et-al-2023}.

\section{Setup}

We first describe the event identification framework, introduced in \cite{N-et-al-2023}, and the two classification models we consider. 

\subsection{Event Identification Framework}\label{eID}
We focus on identifying two classes of events: generation loss (GL) and load loss (LL), denoted by the set $\mathcal{E} \in \{\text{GL, LL}\}$. These events are measured using $M$ PMUs, each of which has access to three channels, namely, voltage magnitude, voltage angle, and frequency, indexed via the set $\mathcal{C}=\{V_m, V_a, F\}$. 
For a given event in $\mathcal{E}$ and channel $c\in \mathcal{C}$, the collected time-series data from $M$ PMUs yields a matrix $x^c\in \R^{M \times N}$, where $N$ is the length of the sample window. Thus, for a given event, the data collected is given by $x = [[x^{V_m}]^T, [x^{V_a}]^T, [x^{F}]^T ]^T \in\R^{|\mathcal{C}| M \times N}$, where $T$ denotes transpose of a matrix/vector. 

In order to evaluate the robustness of this event identification framework, we follow the same feature extraction technique as in \cite{N-et-al-2023} by assuming that the system dynamics can be captured by using modal decomposition to extract a small number $p$ of dominant modes that represent the interacting dynamics of power systems during an event. We refer the reader to \cite{rank1, rank2} for more details on modal decomposition used in this context. These  dynamic modes are defined by their frequency, damping ratio, and residual coefficients that comprise the presence of each mode in a given PMU \cite{mpm, mpm2}. The mode decomposition model is:
\begin{equation}\label{eq:modalrep}
x^c_i(n) = \sum_{k=1}^{p}  R^c_{k,i} \times (Z^c_k)^n  + \epsilon^c_i(n), \quad i  \in \{1, \ldots , M\},  \quad c \in \mathcal{C}
\end{equation}
where $x^c_i(n)$ is the time-series signal for the $i\Th$ PMU and channel $c \in \mathcal{C}$, $R^c_{k,i}$ is the residue for the $k\Th$ mode and $i\Th$ PMU, $Z^c_k = \exp(\lambda^c_k T_s)$ is the $k\Th$ event mode with $\lambda^c_k= \sigma^c_k \pm j \omega^c_k$ and $T_s$ is the sampling period, and $\epsilon^c_i(n)$ is noise. The mode $\lambda^c_k$, defined by $\sigma^c_k$ and $\omega^c_k$, representing the damping ratio and angular frequency of the $k\Th$ mode, respectively. The residue $R^c_{k,i}$ is denoted by its magnitude $|R^c_{k,i}|$ and angle $\theta^c_{k,i}$. The dynamic response to an event is captured by a subset of the system PMUs ($M'< M$) which are chosen based on the highest PMUs' signal energy for a given channel and event. Finally, by extracting the values described above for a given channel and event, we define the feature vector as
\begin{equation}
     X =  \big[\{\omega^c_k\}_{k=1}^{p'},
 \{\sigma^c_k\}_{k=1}^{p'},
    \{|R^c_{k,i}|\}_{k=1}^{p'},
    \{\theta^c_{k,i}\}_{k=1}^{p'} \big]_{i \in \{1, \ldots , M'\},c\in\mathcal{C}}
\end{equation}
Here, we select only the first $p'=p/2$ modes, since typically modes are composed of complex conjugate pairs; by choosing the first $p'$ modes, we keep only one of each conjugate pair.
\par
To compose the overall dataset, we assign event class labels as $y_i=-1$ and $y_i=1$ for LL and GL, respectively. Taking such pairs of event features and their labels, we define the overall dataset as $\mathbf{D} = \{\mathbf{X}_D, \mathbf{Y}_D\}$ where $\mathbf{X}_D = [X_1, ..., X_{n_D}]^T\in \R^{n_D\times d}$, $\mathbf{Y}_D = [y_1, ..., y_{n_D}] \in \R^{n_D }$, and $n_D$ is the total number of events from both classes.

\subsection{Classification Models}

We use logistic regression (LR) and gradient boosting (GB) classification models as the ML models for the evaluation of the framework and design of adversarial attacks. For LR, classification requires computing the probability of event $y_i$ as
\begin{equation}\label{LR_P}
    P(\mathbf{Y} = y_i|X_i,w) = \frac{1}{1+\exp(-y_iw^TX_i)}
\end{equation}
where $w$ is the separating hyperplane between the two classes that would minimize the average classification error over the training data. The optimum estimator is obtained by minimizing the logistic loss as:
\begin{equation}\label{LR_loss}
    w_{\text{LR}} = \arg \min_{w} \sum_{i=1}^{n} \log(1+\exp(-y_iX_i^Tw)).
\end{equation}
Gradient boosting is an ensemble learning algorithm which builds on weak learners, that in our case are decision stumps (single level decision trees thresholded on one feature), each based on a single feature. GB models are trained
with an iterative greedy approach which minimizes error of each new weak learner
by fitting to the residual error made by the previous learned predictors \cite{gb}. The output of the GB model is
\begin{equation}\label{GB}
    F(X)=\sum_{m=1}^{d'} \text{dt}_m(X),\quad \text{where }\text{dt}_m(X)=\begin{cases}v_{1m}, & X_{j_m}\le \text{th}_m\\ v_{2m}, & X_{j_m} > \text{th}_m.\end{cases}
\end{equation}
where $\text{dt}_m$ is the $m^\text{th}$ decision tree with its regression output being $v_{1m}$ or $v_{2m}$ and thresholding the $j_m$ feature at $\text{th}_m$.
%
%
The final GB classifier is obtained by mapping $F(X)$ to the $[0,1]$ range using a sigmoid function and thresholding at $0.5$. 

\section{Threat Models}
In order to evaluate the vulnerability of the event identification framework, we consider two settings: (i) white box; and (ii) gray box. In the white box attack setting, we assume the following: (a) the attacker has full knowledge of the event identification framework, (b) access to all measurements and their corresponding ground truth event label \emph{but} with restricted ability to only tamper with a subset of PMUs, and (c) knowledge of the ML classifier used by the system operator, including all the parameters of the classifier learned by the operator. 

In the gray box attack setting, while assumptions (a) and (b) on the adversarial capabilities still hold, we now assume that the attacker does not know the classification model used by the system operator, but has access to historical data that is not necessarily the same as that used to train the classifier. In either case, our attack algorithm is designed to spoof a specific classifier: in the white box setting, this classifier is the true classifier used by the operator; in the gray box setting, it is a different classifier trained on the adversary's own data.

\begin{figure*}[!t]
\centering
\includegraphics[trim={20 100 20 100 },clip, width=\linewidth]{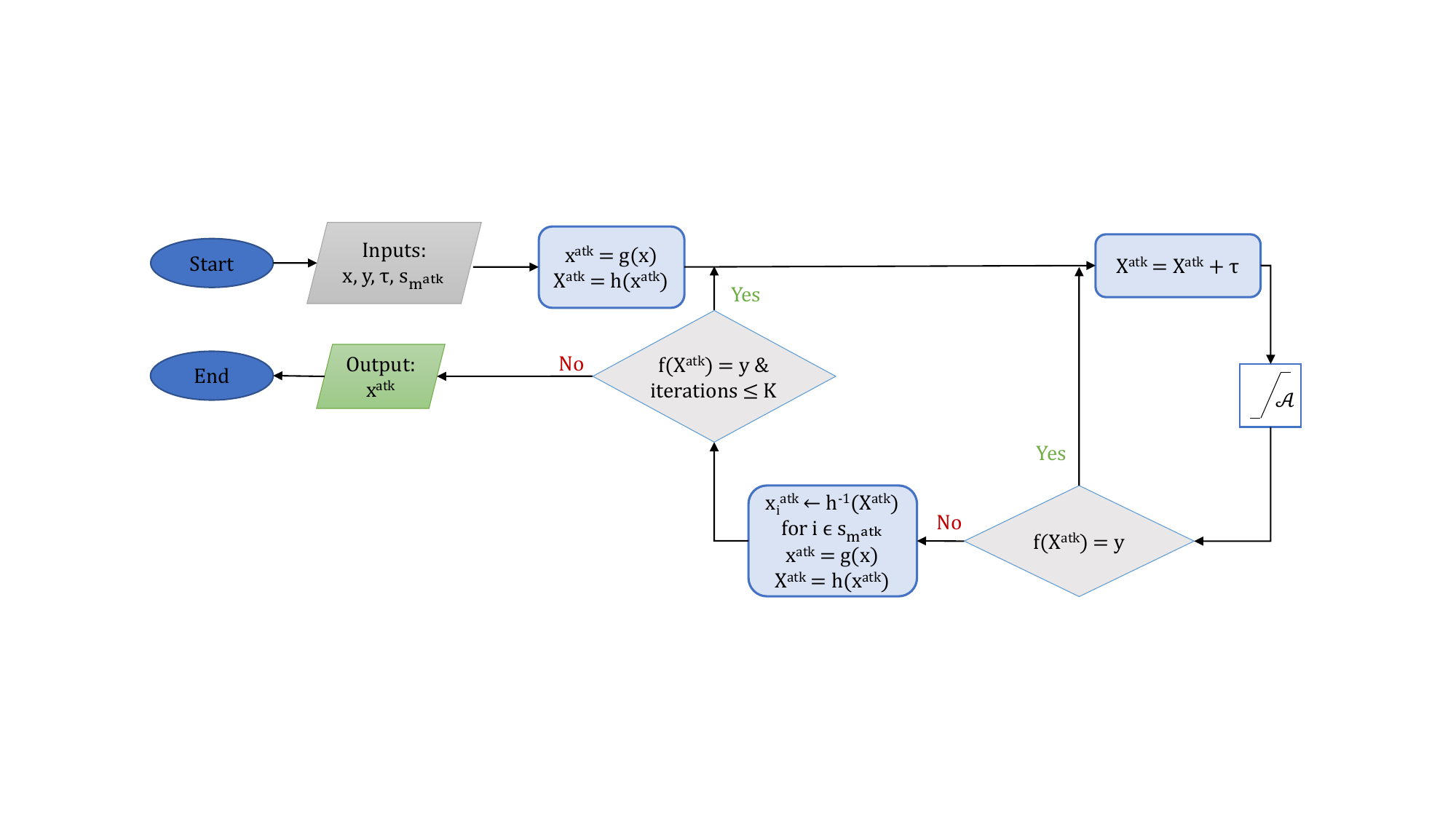}
\caption{\small Attack algorithm using perturbation vector $\tau$ on subset of targeted PMUs $\mathcal{S}_{M^\text{atk}}$ to generate adversarial PMU data $x^{\text{atk}}$. The function $g$ is a signal energy boosting function, $h$ is a modal decomposition conversion function to extract features, $f$ is the classifier, and $\mathcal{A}$ is the feasible set of feature values.}
\label{atkflow}
\end{figure*}

\subsection{Targeted Adversarial Example Generation}
Algorithm \ref{atk} (illustrated in Fig.~\ref{atkflow}) describes how we generate adversarial PMU data. The algorithm utilizes the knowledge of classification models to perturb an incoming feature vector 
such that the direction of the perturbation is chosen to point towards the negative gradient of the classifier. 
The tampered vector of features is then reconstructed to obtain a time domain signal for each of the PMUs tampered by the attacker which then replace the original measurements at these PMUs. 
The resulting collated tampered and untampered event data across all PMUs is passed through the learned classifier for reclassification. This entire procedure is repeated until the classifier fails to classify correctly or when a maximum number of iterations $K$ is reached. 


We explain the steps of Algorithm~\ref{atkflow} as follows. The function $h$ represents the the transform function described in Section \ref{eID}, and $f$ is the classification model used by the attacker (in the white box setting, this is the same as that at the control center; in the gray box setting, it is different). First, we check whether $f(h(x))=y$; that is, whether the event is classified correctly. If not, there is no need to attack it. Next, we start with the untampered time domain data $x$ and boost it so that the PMUs controlled by the adversary are present in the feature vector; this step is represented by the function $g$, which outputs the initial time domain attack vector $x^{\text{atk}}$. In particular, recall from Section~\ref{eID} that only for the $M'$ PMUs with highest energy are the modal residues kept in the feature vector $X$. To ensure that the PMUs controlled by the attacker, denoted by the set $\mathcal{S}^{M_\text{atk}}$, are among these $M'$ PMUs, their energy is boosted by applying $x^{\text{atk}}_i\gets \lambda x^{\text{atk}}_i$ iteratively for all $i \in \mathcal{S}^{M_\text{atk}}$, where $\lambda>1$, until the set $\mathcal{S}^{M_\text{atk}}$ is included in the set of $M'$ PMUs kept in the feature vector.
%

The perturbation vector $\tau$, which is designed based on the classification model $f$ and is meant to be a vector such that changing the feature vector in the direction of $\tau$ will cause the event to be misclassified. The precise details of designing $\tau$ are described in the next subsection. The feature vector $X^{\text{atk}}$ is extracted from $x^{\text{atk}}$ and perturbed by $\tau$ until $f(X^{\text{atk}}) = y'$ where $y'$ is the incorrect event class label. To ensure that the tampered signal remains within reasonable bounds,
the feature classes are restricted to lie within a feasible set $\mathcal{A}$, defined as
\begin{align}
\mathcal{A}&=\{X:|\omega_k^c| \leq \upsilon_{\omega},\ \sigma_k^c > \upsilon_{\sigma},\ \upsilon_{|R|,min} \leq |R_{k,i}^c| \leq \upsilon_{|R|,max},\nonumber
\\&\qquad
\text{for all }k,c,i\}.
\end{align}
where $\upsilon_{\omega}$, $\upsilon_{\sigma}$, and $\upsilon_{|R|,min}$ and $\upsilon_{|R|,max}$ are the bounds for frequency mode, damping mode, and residual amplitude features.
Note that $\theta$ is not restricted since any numerical value of it will be equivalent to a value in $[-\pi, \pi]$ when performing the modal analysis transformation but allows a larger set of feasible and relevant attacks. After perturbing $X^{\text{atk}}$ by $\tau$, it is projected onto $\mathcal{A}$ to ensure it is feasible. 

Once the event features are misclassified in the inner loop, the time domain signals for the compromised PMUs are boosted before replacing the original signal replaces the original signal. 
The resulting tampered time-domain signal is denoted by $x^{\text{atk}}_i$, where $x^{\text{atk}}_i\gets h^{-1}(X^{\text{atk}},i)$, and $h^{-1}$ denotes the inverse of  feature extraction transform that recovers the time domain signal for the $i\Th$ PMU (given by \eqref{eq:modalrep} without the noise term). After reconstructing $x^{\text{atk}}_i$, those time-domain signals are once again boosted via function $g$. Since the feature vector is related to all PMUs, but the attacker can only control a subset, the resulting time-domain attack vector $x^{\text{atk}}$ will not exactly match the feature vector $X^{\text{atk}}$. Thus, $X^{\text{atk}}$ is recomputed using feature extraction, and the loop repeats.


\subsection{Designing $\tau$}\label{fatk}

We describe how to find the perturbation vector $\tau$ used to design attacks in Algorithm~\ref{atkflow} based on the classifier $f\in \{f_{LR}, f_{GB}\}$.
For the LR classifier, we designate the separating hyperplane by its weight vector $w^{LR}$, as in \eqref{LR_P}. Thus, we can misclassify an event by perturbing its values towards the hyperplane. To realize this, we let $\tau = -y_i \,\eta\, w^{LR}$ for event $i$, where $\eta \in \R$ is a  step size chosen sufficiently small to avoid perturbing event features too much. 

For GB, recall that the classifier is composed of a sum of $d'$ decision trees, given by equation \eqref{GB}. The $m\Th$ decision tree $\text{dt}_m$ is applied to the feature $j_m$ and is described by its two values $v_{1m},v_{2m}$ and the threshold $\text{th}_m$. A crude approximation of the gradient of GB model can be written as:
\vspace{-.1in}
\begin{equation}\label{GB_w}
    w^{GB}_m= v_{1m} - v_{2m}
\end{equation}
where $w^{GB}_m$ defines the weight and direction of the approximated gradient from $\text{dt}_m$. Thus, if $w^{GB}_m$ is positive, we increase the value of the $j_m^\text{th}$ feature if $y_i=1$ and decrease it if $y_i=-1$. (Vice versa if $w^{GB}_m$ is negative.) By doing so, we are forcing regression trees to output the less favorable value, leading to misclassification. In cases where multiple decision trees act on the same feature, and potentially have opposite signs of $w^{GB}_m$, the magnitude of $w^{GB}_m$ for a given decision tree will signify its importance on the overall output of GB classification. Now we define the perturbation vector $\tau$ as the $d$-dimensional vector whose $j^\text{th}$ entry, for $j=1, \ldots, d$, is
\begin{equation}\label{GB_tau}
    \tau_j=\eta\, y_i \sum_{m:j_m=j} w^{GB}_{m}.
\end{equation}
By \eqref{GB_tau}, If the same feature is used in multiple trees, then this feature will be adjusted in proportion to the tree importance described in \eqref{GB_w}.

\par 

\begin{algorithm}
\caption{Targeted Adversarial Example Generation}\label{atk}
\begin{algorithmic}
    \State{\textbf{Input: }$x, y$: untampered PMU data and true label}
    \State{$\quad \qquad f$: Classification model}
    \State{$\quad \qquad h$: Feature extraction transform}
    \State{$\quad \qquad g$: Signal energy boosting function}
    \State{$\quad \qquad \tau$: Perturbation vector}
    \State{$\quad \qquad \mathcal{A}$: Feasible feature set}
    \State{$\quad \qquad \mathcal{S}^{M_\text{atk}}$: Set of PMUs controlled by attacker}
    \State{\textbf{If: $f(h(x)) = y$ do}}
    \State{\textbf{Initialize: } $x^{\text{atk}} \gets g(x,\mathcal{S}^{M_\text{atk}})$}
    \State{$\quad \qquad \qquad X_{\text{atk}} \gets h(x^{\text{atk}})$}
    \While{$f(X^{\text{atk}}) = y$ and iterations $\leq K$}
        \While{$f(X^{\text{atk}}) = y$}
            \State $X^{\text{atk}} \gets X^{\text{atk}} +  \tau$ 
            \State Project $X^{\text{atk}}$ into $\mathcal{A}$
        \EndWhile
        \State{\textbf{for all $i \in \mathcal{S}^{M_\text{atk}}$ do}}
            \State{$\qquad  x^{\text{atk}}_i \gets h^{-1}(X^{\text{atk}},i)$}
        \State{\textbf{end for}}
        \State $x^{\text{atk}} \gets g(x,\mathcal{S}^{M_\text{atk}})$
        \State $X^{\text{atk}} \gets h(x^{\text{atk}})$
    \EndWhile
    \State{\textbf{Return:}  $x^{\text{atk}}$}
\end{algorithmic}
\end{algorithm}\par

\section{Numerical Results}
\subsection{Dataset}
The synthetic South Carolina 500-bus grid, consisting of 90 generators, 466 branches, and 206 loads \cite{500-1}, is used to generate synthetic generation loss and load loss events. A dynamic model of the system on PSS/E is used to generate event data by running dynamic simulations for 11 seconds at a sampling rate of 30Hz. The event is applied after 1 second to ensure the system has reached steady-state. Data is collected from PMUs distributed on the largest $M=95$ generator and load buses  of the network (largest in terms of net generation or load). The GL events are generated by disconnecting the largest 50 generators, one per simulation run. For each such generator, 15 different loading scenarios are considered where the overall system loading varies between 90\% to 100\% of the net load. This is done by varying each individual load in the system randomly within its operational limits. Through this process, we obtain a total of 750 GL events. We create the LL events in a similar manner (i.e., disconnecting the largest 75 loads, one at a time, at 10 different loading scenarios varying between 90\% to 100\%). Thus the complete dataset has a total of $n_D = 1500$ event samples collected from voltage magnitude, voltage angle, and frequency channels of $M=95$ PMUs.

In order to train the ML classification models, the dataset is split into three sets: 20\% testing set and training sets for LR and GB each consisting of 40\% of the dataset. Each set is assured to be nearly balanced across the two classes of events.

\subsection{Evaluation of Base Cases}
\begin{figure}[!t]
\centering
\includegraphics[trim={6 6 6 6 },clip, width=\linewidth]{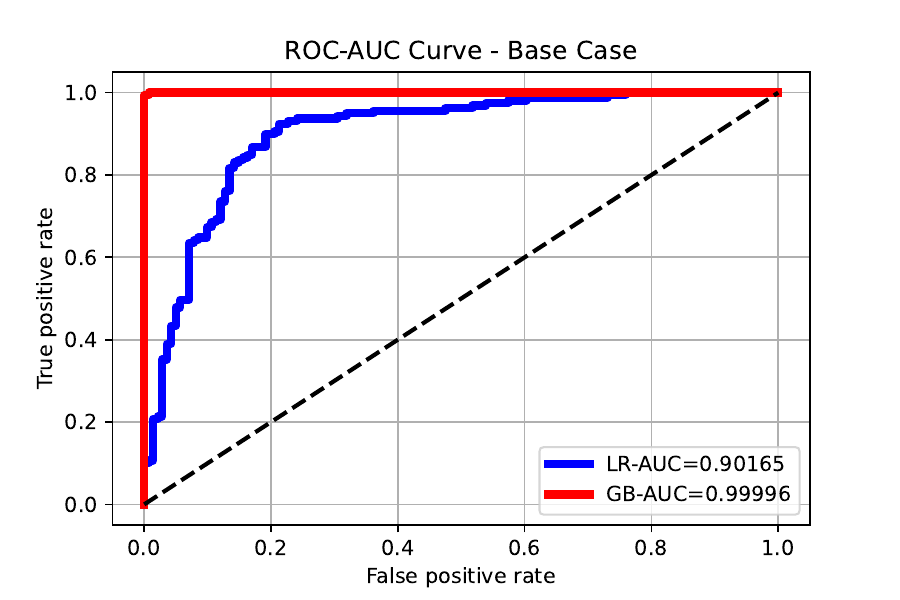}
\caption{\small Base case (untampered) performance of LR and GB classification models evaluated on the testing set.}
\label{basecase}
\end{figure}

Figure \ref{basecase} shows the base case performance of both models. Note that the LR and GB classifiers are trained on their respective untampered training set and evaluated on untampered testing set. The resulting test accuracy is shown in the figure. To this end, we use receiver operating characteristic area under the curve (ROC-AUC) as the accuracy metric to evaluate the performance of the base and tampered models. The base models are able to identify unseen data with high accuracy with the GB model approaching 100\% accuracy and surpassing the performance of LR. 

\subsection{Generation and Evaluation of Adversarial Examples} \label{taes}

\begin{figure}[th]
\centering
\includegraphics[trim={7 6 7 6 },clip, width=\linewidth]{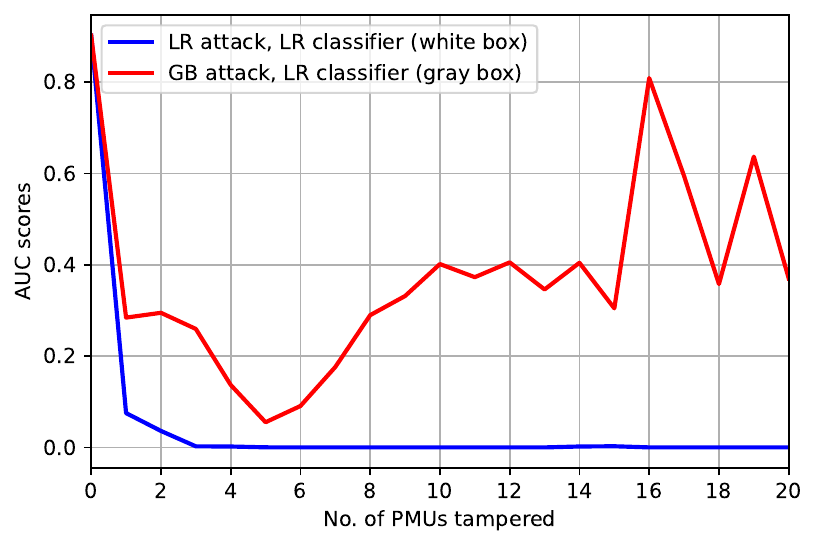}
\caption{\small AUC scores as a function of the number of tampered PMUs for white (blue curve) and gray (red curve) box attacks for the logistic regression (LR) classifier.}
\label{lrhybrid}
\end{figure}

\begin{figure}[th]
\centering
\includegraphics[trim={7 6 7 6 },clip, width=\linewidth]{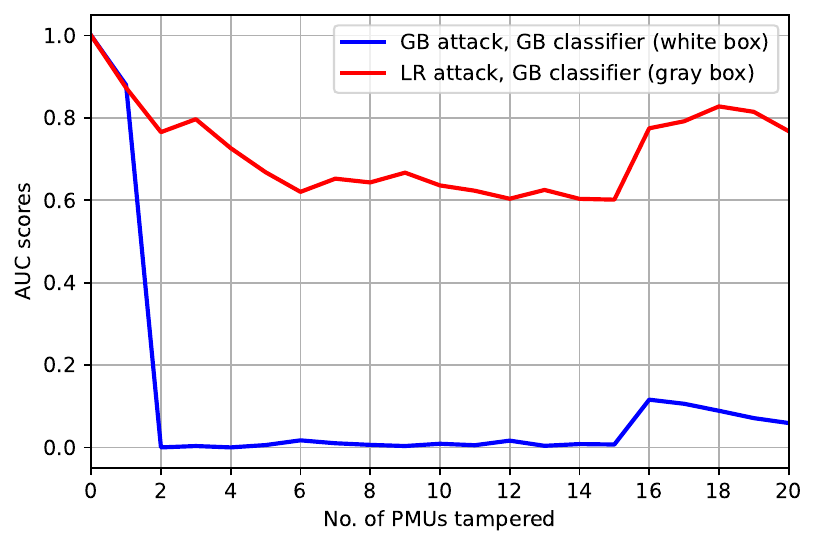}
\caption{\small AUC scores as a function of the number of tampered PMUs for white (red curve) and gray (blue curve) box attacks for the gradient boosting (GB) classifier.}
\label{gbhybrid}
\end{figure}

As a first step towards evaluating the white box and gray box attack algorithms, we generate the tampered events as outlined earlier. The average AUC scores are plotted in Figures \ref{lrhybrid} and~\ref{gbhybrid}. In short, we iterate over the original events from the testing set as input to the attack algorithms and choose the feasible set bounds as 
\begin{equation}
\upsilon_{\omega}=2\omega_0,\ \upsilon_{\sigma}=0,\ \upsilon_{|R|,min}=0.8|R_0|,\  \upsilon_{|R|,max}=2|R_0| 
\end{equation}
where $\omega_0$ and $|R_0|$ are the untampered values of those features for a given event. 

To evaluate the impact of attacks on different numbers of PMUs, we choose 10 random sets $\mathcal{S}_{\text{atk}}$, each consisting of $M'=20$ PMUs. Denote $\mathcal{S}_{M^\text{atk}}$ 
as the set consisting of the first  $M^{\text{atk}}$ 
PMUs in $\mathcal{S}_{\text{atk}}$, 
where $M^{\text{atk}}$  
varies from $1$ to $20$. We then evaluate the attack on $\mathcal{S}_{M^\text{atk}}$ for each $M^{\text{atk}}$.  
Figures~\ref{lrhybrid} and~\ref{gbhybrid} show the average AUC as a function of $M^{\text{atk}}$. 

We evaluate white and gray box attacks as follows.
Let $f\in\{\text{LR},\text{GB}\}$ be the classification model used in the attack algorithm.  In the white box setup, $f$ is used both in the attack algorithm and as the classifier applied to the generated attack data. In the gray box setup, $f$ is only used in the attack algorithm; and  the classification model in $\{\text{LR},\text{GB}\}$ other than $f$ is used as the classifier.
In other words, we run the attack algorithm using the knowledge of both classification models (LR and GB) and evaluate the output from each case using both classifiers. 

For white box attacks on both LR and GB classifiers, the accuracy of the classifiers drops to close to $0\%$ even when only 2 or 3 PMUs are attacked. In the gray box setup, attacks show a significant decrease in accuracy; however, they are less successful than white box attacks. This is expected as these attacks are designed to target different models. Moreover, we observe that GB models are more robust against gray box attacks compared to LR. Finally, gray box attacks on LR show a fluctuating behavior as the number of tampered PMUs increases. This is likely a result of the attack being tailored for GB, leading to unpredictable effects on the LR classifier. That is, an attack with the power to control more PMUs will not necessarily be more effective, since it may be pushing in the wrong direction.

\section{Conclusion}
ML-based event classification techniques can enhance situational awareness, especially with increasing DER penetration and their need for fast dynamic monitoring and response. We have shown that white box attacks for both LR and GB classifiers are highly successful, reducing the AUC score significantly with only a few PMUs tampered. On the other hand, gray box attacks cause a relatively modest reduction in AUC scores with GB being more robust. With our attack showing vulnerabilities in ML classifiers, future work will include developing classifiers to be more robust against attacks, as well as classifiers that are designed to distinguish attacks from legitimate events.
\balance
\bibliographystyle{IEEEtran}
\bibliography{References.bib}

\end{document}